\documentclass[%
 aip,
 jmp,%
 amsmath,amssymb,
 reprint,%
]{revtex4-1}

\usepackage{graphicx}
\usepackage{dcolumn}
\usepackage{bm}

\begin{document}


\title{Direct observation of the M2 phase with its Mott transition in a VO$_2$ film}

\author{Hoon Kim}
\affiliation{Center for Artificial Low Dimensional Electronic Systems, Institute for Basic Science (IBS), Pohang 37673, Republic of Korea}
\affiliation{Department of Physics, Pohang University of Science and Technology (POSTECH), Pohang 37673, Republic of Korea}

\author{Tetiana V. Slusar}
\affiliation{Electronics and Telecommunications Research Institute, 218, Gajeong-ro, Yuseong-gu, Daejeon, 34129, Korea}

\author{Dirk Wulferding}
\affiliation{Center for Artificial Low Dimensional Electronic Systems, Institute for Basic Science (IBS), Pohang 37673, Republic of Korea}

\author{Ilkyu Yang}
\affiliation{Center for Artificial Low Dimensional Electronic Systems, Institute for Basic Science (IBS), Pohang 37673, Republic of Korea}

\author{Jin-Cheol Cho}
\affiliation{Electronics and Telecommunications Research Institute, 218, Gajeong-ro, Yuseong-gu, Daejeon, 34129, Korea}

\author{Minkyung Lee}
\affiliation{Center for Artificial Low Dimensional Electronic Systems, Institute for Basic Science (IBS), Pohang 37673, Republic of Korea}
\affiliation{Department of Chemistry, Pohang University of Science and Technology (POSTECH), Pohang 37673, Republic of Korea}

\author{Hee Cheul Choi}
\affiliation{Center for Artificial Low Dimensional Electronic Systems, Institute for Basic Science (IBS), Pohang 37673, Republic of Korea}
\affiliation{Department of Chemistry, Pohang University of Science and Technology (POSTECH), Pohang 37673, Republic of Korea}

\author{Yoon Hee Jeong}
\affiliation{Department of Physics, Pohang University of Science and Technology (POSTECH), Pohang 37673, Republic of Korea}

\author{Hyun-Tak Kim}
\affiliation{Electronics and Telecommunications Research Institute, 218, Gajeong-ro, Yuseong-gu, Daejeon, 34129, Korea}
\affiliation{School of Advanced Device Technology, Korea University of Science \& Technology, Daejeon 34113, South Korea}

\author{Jeehoon Kim}
\email[]{Corresponding author: jeehoon@postech.ac.kr}
\affiliation{Center for Artificial Low Dimensional Electronic Systems, Institute for Basic Science (IBS), Pohang 37673, Republic of Korea}
\affiliation{Department of Physics, Pohang University of Science and Technology (POSTECH), Pohang 37673, Republic of Korea}

\date{\today}

\begin{abstract}

In VO$_2$, the explicit origin of the insulator-to-metal transition is still disputable between Peierls and Mott insulators. Along with the controversy, its second monoclinic (M2) phase has received considerable attention due to the presence of electron correlation in undimerized vanadium ions. However, the origin of the M2 phase is still obscure. Here, we study a granular VO$_2$ film using conductive atomic force microscopy and Raman scattering. Upon the structural transition from monoclinic to rutile, we observe directly  an intermediate state showing the coexistence of monoclinic M1 and M2 phases. The conductivity near the grain boundary in this regime is six times larger than that of the grain core, producing a donut-like landscape. Our results reveal an intra-grain percolation process, indicating that VO$_2$ with the M2 phase is a Mott insulator.

\end{abstract}

\maketitle
Among the Magn\'eli phases of vanadium oxides, VO$_2$ shows an insulator-to-metal transition (IMT) at 340 K, which is remarkably close to room temperature \cite{Kachi}. On top of it, the tunability of the transition involves various aspects of defects \cite{Byczuk, Fan, Appavoo}, gate voltage \cite{Stefanovich}, and photon excitation \cite{Morrison}. In addition, its sensitivity to lattice modifications has triggered several investigations such as doping \cite{Pouget_Cr}, strain \cite{Mengkun, Okimura_RTM2, Cao, Atkin, Okimura_T}, and pressure \cite{Pouget_Pressure}. Owing to the applicability and flexibility of the transition, VO$_2$ has been widely examined to uncover its transition nature.

From the electronic point of view, a Mott insulator due to on-site Coulomb interaction experiences an IMT, when the Hubbard gap is reduced and is overcome by the attractive electron-hole pair energy \cite{Mott, Kim_Hole}. At the crossover, a large number of free charge carriers form discontinuously, meeting Mott's criteria of the critical carrier density \cite{Mott, Kim_Hole, Edwards, Stefanovich}. On the other hand, the IMT in VO$_2$ is accompanied by a structural phase transition (SPT) from a monoclinic (M1) to a rutile (R) phase. The driving force initiating the IMT remains unclear since both occur concomitantly. Therefore, two mechanisms are competitively discussed at present: (1) a Peierls instability, {\it i.e.}, a dimerization of vanadium atoms, and (2) a Mott IMT. The former is supported by band-structure calculations, considering a homopolar V-V bond to split $d_{\parallel}$ orbitals \cite{Goodenough}, and by DFT-LDA calculations \cite{Wentzcovitch}. It is further corroborated by a simultaneous observation of both SPT and IMT within the internal thermalization time of a pump-probe method \cite{Cavalleri_1} and the measurement of a delayed structural response time upon a pulsed excitation \cite{Cavalleri_2}. The latter mechanism, which induces a gap by repulsive electron-electron correlation, is mainly supported by experimental evidence. In the context of a Mott IMT, examples include a diverging effective mass at the crossover \cite{Qazilbash}, an estimation of the temperature difference between SPT and IMT \cite{Kim_Coherent}, a metallic monoclinic phase \cite{Morrison, Kim_Coherent, Kim_Schottky, Tao, Laverock}, and an IMT triggered by electron injection \cite{Stefanovich, Wegkamp}. The conflicting results also have implications on the cooperativity between the two mechanisms, but the conclusive origin of the IMT is still under debate \cite{Kim_Schottky}.

The application of uniaxial pressure leads to the formation of a second monoclinic (M2) phase in VO$_2$ between the M1 and R phases \cite{Pouget_Pressure, Rice}. Only half of the vanadium ions participate in the dimerization (site 1) in the M2 phase, which is more insulating than the M1 phase  \cite{Cao, Zylbersztejn}.  
The undimerized vanadium ions of M2 (site 2) are reported to have strong electron correlations \cite{Pouget_Cr}. In addition, the emergence of the M2 phase as an intermediate state between the M1 and R phases was proposed to be universal \cite{Atkin}. Furthermore, the M1 phase can be interpreted as a superposition of two M2 phases \cite{Pouget_Pressure}. These observations underline the importance of the M2 phase to clarify the IMT origin in VO$_2$, though the exact condition for the M2 formation is unclear. Compressive stress along the [011]$_{M1}$ direction seems to favor the M2 phase in single crystals \cite{Pouget_Pressure}, micro- and nanobeams \cite{Cao, Atkin, Park, S_Zhang, Guo}. In contrast, granular films require tensile stress in the same direction \cite{Okimura_RTM2, Zhang, Okimura_Selective}. The stabilization mechanism \cite{Pouget_Cr, Zylbersztejn} and the origin of high resistivity \cite{Cao, Zylbersztejn} in the M2 phase are still evasive. However, simultaneous Raman mapping and resistivity dependencies of VO$_2$ nanobeams have revealed two distinct thermally activated behaviors for the M1 and the M2 phase \cite{S_Zhang}. In this case the activation energy of the M2 phase follows a Mott-Hubbard gap forecasting its inherency, but the emergence of the M2 phase in real space was not resolved on granular films.

As a real space probe, scanning near-field infrared microscopy was able to map simultaneously the metallicity and structural distribution in VO$_2$ \cite{Mengkun, Qazilbash, Frenzel, Qazilbash_Inversion}. However, in granular films, its limited resolution prohibits an investigation of the intra-grain properties. In contrast, conductive atomic force microscopy (C-AFM) offers a high resolution \cite{Jeehoon} and provides detailed insight into how the local environment can affect transitions. In this study, we combine temperature-dependent C-AFM and Raman scattering to investigate locally the electronic as well as the structural properties of the M1 and M2 phases in VO$_2$ films. Our choice of a nanogranular film allows a microscopic study of VO$_2$ with a modulated strain varying locally between grain core and grain boundary. As a result, we directly observe the emergence of the M2 phase within a single grain in the matrix of the M1 phase and its intra-grain percolation, which is a fingerprint for a Mott-driven IMT.

A 100-nm-thick VO$_2$ film was grown onto a p-type Si substrate by using the pulsed laser deposition technique \cite{Slusar}. C-AFM was performed using a home-built heating stage and commercially available tips \cite{C2}. Unpolarized Raman scattering experiments have been performed in the backscattering geometry with the excitation line $\lambda = 532$ nm by using a micro-Raman spectrometer (WITEC Alpha 300R Raman spectroscope). The laser power was set to 500 $\mu$W to minimize local heating effects.

\begin{figure}
\includegraphics[width=8.5cm]{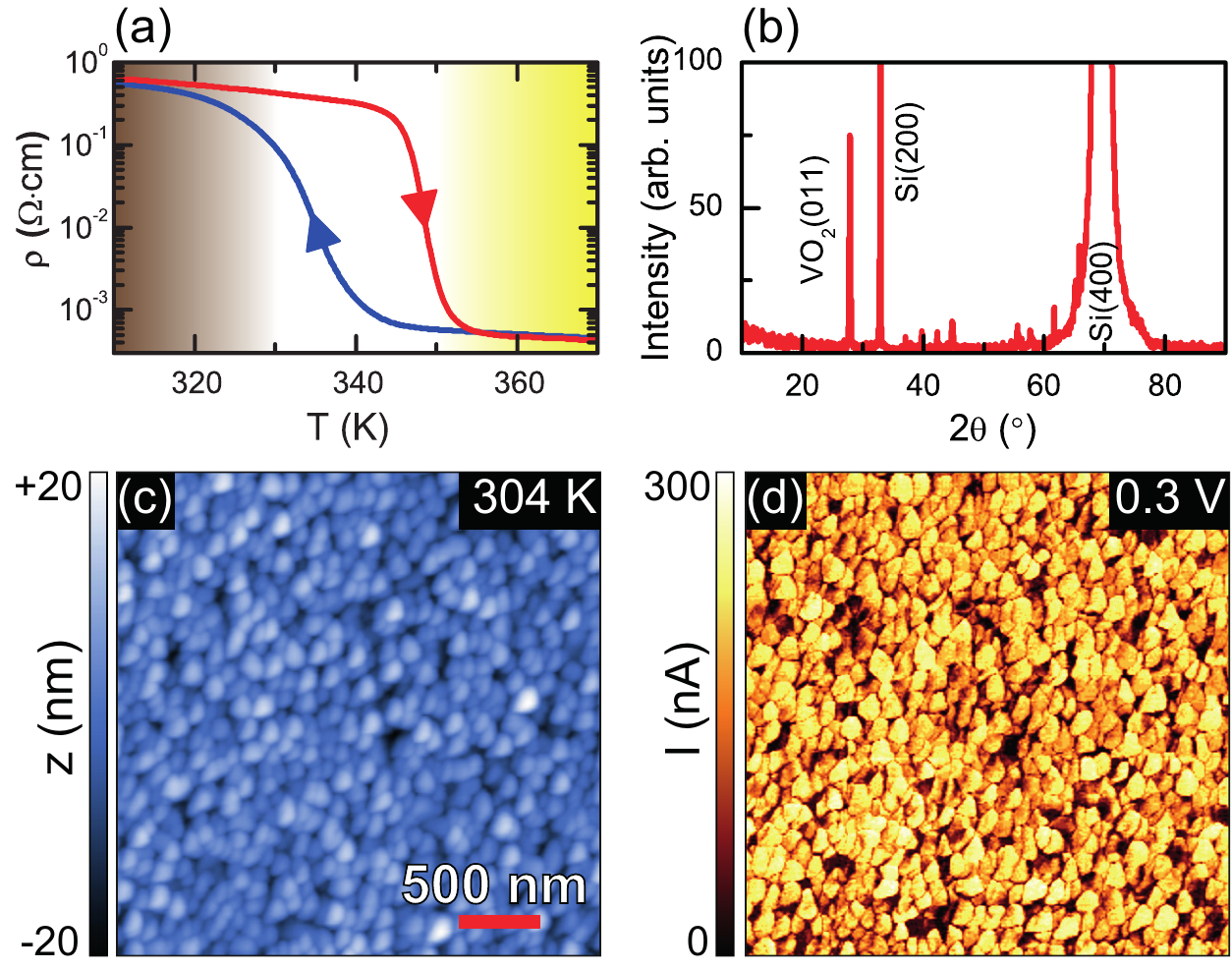}
\caption{(a) Temperature-dependent resistivity curve of the VO$_2$ film. (b) XRD data of the film measured at room temperature. Simultaneously obtained topography (c) and conductance map (d) at $T=304$ K using C-AFM.}
\end{figure}

The onset transition temperature $T_{\mathrm{MI}}$ of our thin film is 344 K in heating curve, as shown in Fig. 1(a), and the points where the curve slope is at its steepest indicate the transition temperatures upon heating and cooling are 349 K and 334.5 K, respectively. The noticeable enhancement of $T_{\mathrm{MI}}$ in comparison to previously reported values \cite{Lu, Kittiwatanakul} might indicate a substantial emergence of the M2 phase \cite{Park}, similar to the previously reported increased $T_{\mathrm{MI}}$ values through tensile strain or epitaxial stress \cite{Kittiwatanakul, Muraoka}. A resistivity drop by three orders of magnitude is comparable to that observed in previous reports \cite{Mengkun, Slusar}. The $\rho$-$T$ hysteresis curve is asymmetric. A film with a range of grain sizes produces hysteresis loops with various widths and centers and thereby a globally asymmetric hysteresis loop \cite{Klimov}. The asymmetric hysteresis loop in our film is narrow at high temperature, indicating the abundance of relatively large grains. Further evidence for the large grain size is found in the topography and conductance map in Figs. 1(c) and 1(d), respectively, obtained simultaneously near room temperature. We find that the topography in Fig. 1(c) has a peak-to-peak value of 40 nm and a root mean square surface roughness of 5.6 nm. A small bias voltage was maintained in the conductance map to avoid local heating effects and electron injection. The average grain size of our film is approximately 200 nm [see Fig. 1(c)], which is larger than that of previous reports ($\sim$ 100 nm) \cite{Jeehoon}.

Since grain boundaries exert tensile strain through inter-grain attraction \cite{Doljack, Janssen_tensile}, the larger grains with a lower density of boundaries can be dominated by compressive strain \cite{Okimura_RTM2, Janssen_compressive}. In granular films, the M2 phase is stabilized when the compressive strain is applied in the direction perpendicular to [011]$_{M1}$ \cite{Okimura_RTM2, Zhang}. The room temperature XRD data of our film reveals a sharp peak corresponding to [011]$_{M1}$ of VO$_2$ [Fig. 1(b)], hence, compressive strain causes the emergence of the M2 phase.

\begin{figure}
\includegraphics[width=8.5cm]{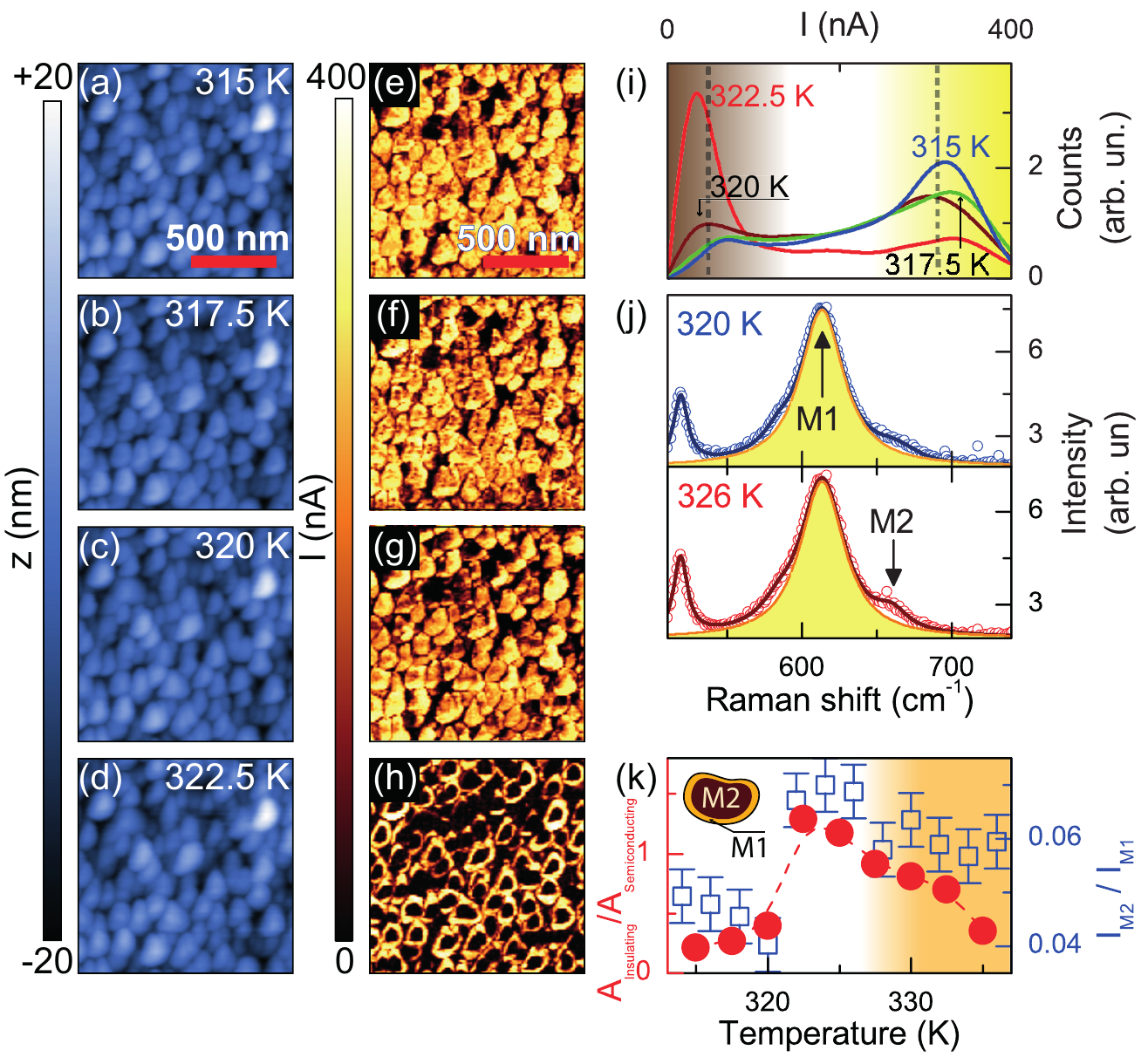}
\caption{Simultaneously obtained topography (a)--(d) and conductance maps (e)--(h) with increasing temperature. A tip-sample bias of 0.2 V was applied. In particular, Fig. 2(h) shows a donut-like conductivity distribution, which is an important discovery of this study. (i) Histograms of the conductance obtained from (e)--(h). The shaded background marks two distinct phases, corresponding to the black and yellow regions in (h). (j) Temperature-dependent Raman spectra focusing on the spectral region of phonon modes related to the M1 and M2 phases. (k) The area ratio between the insulating and the semiconducting phases in C-AFM (red bullets), together with the intensity ratio of the M2 and M1 phonon modes in Raman scattering (blue squares) plotted as a function of temperature.}
\end{figure}

The temperature-dependent conductance maps and simultaneously obtained topography are shown in Figs. 2(e)--(h) and Figs. 2(a)--(d), respectively. In the conductance maps, the semiconducting region becomes more inhomogeneous with the growth of the insulating region. Finally, the insulating phase abruptly emerges and fully occupies the grains cores, shaping the previously homogeneous grains into donuts [see Fig. 2(h)]. This is our key observation. The donut-shape transition occurs far below $T_{\mathrm{MI}}$, indicating the manifestation of another transition within the monoclinic phase. This phenomenon is related to the formation of the M2 phase within the grain, as discussed below. The histograms extracted from the conductance maps show a distinct bimodal behavior between the two phases, marked by the brown and yellow shaded regions in Fig. 2(i). According to the strain-temperature phase diagram in Ref. [11], the observed bimodal behavior is a characteristic feature for the M1 and M2 phases. Above $T$ = 305 K, the transient triclinic phase disappears; consequently, both M1 and M2 phases emerge separately. Therefore, the bright and dark phases in Figs. 2(e)--(h) are related to the M1 and M2 phases.

In order to investigate the structural component upon the phase transition, we performed Raman scattering as a function of temperature. The laser beam was focused onto a spot with 1-$\mu$m diameter, averaging over several grains. Two distinct phonon modes corresponding to the M1 and M2 phases appear in the spectral range of 600 -- 700 cm$^{-1}$ [Fig. 2(j)]. The phonon frequencies are in accordance with previous studies \cite{Atkin, Okimura_T, Zhang}. We observe an increase of the M2 phonon intensity at 322 K, which bears similarity with the drastic emergence of the insulating phase in the conductance maps. Fig. 2(k) shows the intensity ratio of M2 : M1 obtained from Raman scattering and C-AFM, exhibiting good agreement. A clear transition between 320 K and 322 K is evident from both measurements. Therefore, the insulating phase observed in the conductance maps corresponds to the M2 phase.
The tip pressure on the sample surface is estimated to be $\sim$700 bar, which is greater than the reported value of 300 bar, to initiate the M2 phase in a single-crystal VO$_2$ \cite{Pouget_Pressure}. However, the conductivity map is not affected when the pressure is reduced to $\sim$350 bar. Thus, the tip pressure is negligible compared to the film's innate strain.
It is worth noting that C-AFM is a strictly surface sensitive technique, while Raman spectroscopy can have a probing depth of several hundred nanometers into the material. Therefore, the M2 : M1 ratios obtained from both methods are vastly different, yet, the commonly observed transition around 322 K strongly supports the emergence of the M2 phase.

The phase separation between the grain boundary and core suggests possible filamentary conduction along grain boundaries, and explains the frequent absence of the M2 signature in bulk resistivity measurements \cite{Zylbersztejn}. A recent transport study on a strained microbeam of VO$_2$ reported a resistivity ratio M2/M1 of 3.\cite{Cao} According to the histograms in Fig. 2(i), we obtained a resistivity ratio M2/M1 of 6. Note that the filamentary conduction does not affect our local conductivity measurements. This would explain the reduced resistivity ratio in bulk measurements where the filamentary conduction prevails.

\begin{figure}
\includegraphics[width=8.5cm]{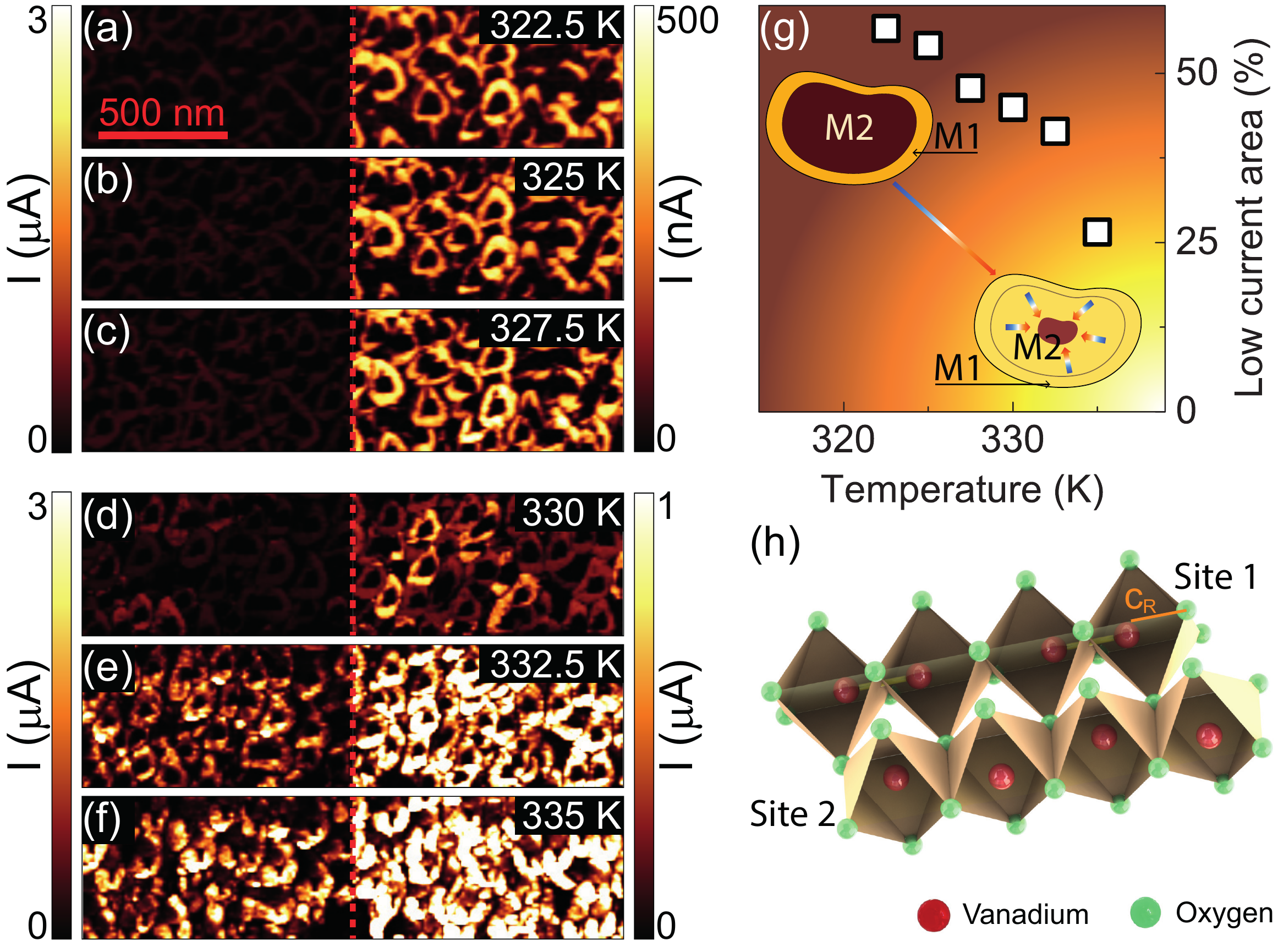}
\caption{(a)--(f) Temperature-dependent conductance maps up to 335 K. The contrast on the left-hand side is adjusted to a global current scale (0--3 $\mu$A), and the right-hand side is rescaled to accommodate minor changes in conductivity for individual maps. (g) The fraction of the low-current area as a function of temperature. Its steady demise represents the progress of intra-grain percolation at the expense of the insulating M2 region. The inset shows a schematic diagram of the intra-grain percolation during a gradual temperature rise. (h) A schematic illustration showing two vanadium sites of the M2 phase. Owing to a thermal drift during the measurements, the images in Figs. 3(a)--(c) and 3(d)--(f) were obtained at two adjacent regions. 
The tip-sample bias was reduced from 0.2 V to 0.1 V in Figs. 3(d)--(f) owing to the predominant increase of conductivity.
}
\end{figure}

According to Mott's criteria, the carrier density of a Mott insulator abruptly increases to the critical value $n_c$ through a first-order transition, when the Hubbard gap diminishes and the electron-hole attraction surmounts it \cite{Mott,  Edwards, Stefanovich}. The extrinsic carriers at half-filling can suppress this Hubbard gap since they effectively decrease the Coulomb repulsion energy for hopping \cite{Kim_Schottky}. The density of charge carriers for initiating the IMT is represented as Kim's criterion, $\Delta \rho ' = 0.018 \%$ \cite{Kim_Hole, Kim_Schottky}, where $\Delta \rho '$ is the ratio of impurities to half-filling carriers. Therefore, a Mott insulator becomes a metal when the concentration of mobile charge carriers exceeds a certain level. In a VO$_2$ film, the impurity bands at grain boundaries contribute to the IMT with excessive charge carriers resulting from oxygen deficiencies. Therefore, in a Mott-driven IMT, the transition evolves from the grain boundary toward the core. While Raman spectroscopy maintains a rather stable M2 : M1 ratio above the M1-M2 transition [see shaded region in Fig. 2(k)], the edge of the insulating M2 phase gradually acquires a semiconducting state, as shown in Figs. 3(a)--(f). Fig. 3(g) plots the portion of the low-current area with respect to temperature, as obtained from Figs. 3(a)--(f), indicating that the intra-grain percolation develops as temperature increases. The observed intra-grain percolation is interpreted as a Mott-type IMT of the M2 phase.

The formation of the M2 phase at the grain core highlights the role of oxygen. It weighs on the interpretation of the M1-M2 transition as a competition between V-O $\pi$-bonding and V-V bonds. When the energy gain resulting from the stabilization of the apical V-O $\pi$-bonds at site 1 exceeds the cost for melting the V-V dimerization at site 2, the M2 phase can emerge, yielding different structures at two vanadium sites, as shown in Fig. 3(h). It is supported by a previous report of a blueshift of the V-O bond energy in Raman spectroscopy through the M1-M2 transition \cite{Atkin, Okimura_T}. This perspective sheds light on the large resistivity ratio between the M2 and M1 phase. The $\pi$-antibonding level of VO$_2$ resides inside a correlation gap between half-filled Hubbard bands from the $d_{\parallel}$ orbital \cite{Zylbersztejn, Goodenough}. When the $\pi$-bonding orbital is stabilized, the energy level of the $\pi$-antibonding orbital is lifted from the lower Hubbard band. This further strengthens gap opening, leading to narrower $d_{\parallel}$ bands by reducing screening and hybridization \cite{Zylbersztejn}. This narrowing of the $d_{\parallel}$ band was observed in photoelectron spectroscopy \cite{Okimura_T}. Thus, the enhancement of the gap suggests an increase in the resistivity of the M2 phase.

In conclusion, we observed the emergence of the M2 phase in a single nanograin of VO$_2$ with a donut-like conductivity distribution. Our observation reveals intra-grain percolation upon the IMT, spreading conductivity from the grain boundaries toward the cores. These results suggest that the M2 phase, involving two different vanadium sites of dimerized and undimerized ions, fully undergoes a Mott-type IMT.

\begin{acknowledgments}
This work was supported by the Institute for Basic Science (IBS), Grant No. IBS-R014-D1, and a creative project at the Electronics and Telecommunications Research Institute (ETRI). YHJ was supported by Center for Topological Matter at POSTECH (2011-0030786).
\end{acknowledgments}


\begin{thebibliography}{99}
\bibitem{C2}
A commercially available Pt-Ir coated tip from AppNano (ANSCM-PT-20) with a spring constant of $k=3$ N/m was used to obtain conductance maps.
\end{thebibliography}
\end{document}